\documentclass{PoS}

\title{Precision $B_c$ and $B_s$ mass calculations}

\ShortTitle{Precision $M_{B_c}$ and $M_{B_s}$}

\author{HPQCD Collaboration}

\author{\speaker{E. B. Gregory}, C. T. H. Davies, K. Y. Wong,\\
        Department Physics \& Astronomy, University of Glasgow, Glasgow, 
G12 8QQ, UK\\
        E-mail: \email{e.gregory@physics.gla.ac.uk}}

\author{E. Follana\\
Departamento de F\'{i}sica Te\'{o}rica, Universidad de Zaragoza, Zaragoza, Spain}

\author{E. Gamiz\\
Physics Department, University of Illinois, Urbana, IL 61801, USA}

\author{G. P. Lepage\\
        Laboratory of Elementary Particle Physics, Cornell University, Ithaca, 
NY 14853, USA\\}
\author{J. Shigemitsu\\
Department of Physics, The Ohio State University, Columbus, OH 43210, USA}

\abstract{We give improved results for B meson masses using NRQCD $b$ quarks 
and HISQ light valence quarks for a range of lattice spacings and sea quark 
masses enabling controlled extrapolation to the physical point.}

\FullConference{The XXVI International Symposium on Lattice Field Theory\\
		 July 14-19 2008\\
		 Williamsburg, Virginia, USA}

\newcounter{saveeqn}
\newcommand{\romeqn}{\setcounter{saveeqn}{\value{equation}}%
  \renewcommand{\theequation}{\Roman{equation}}
}
\newcommand{\reseteqn}{\setcounter{equation}{\value{saveeqn}}%
  \renewcommand{\theequation}{\arabic{equation}}}

\begin{document}

\section{Introduction}
The precise calculation of the charm and bottom spectra is an important 
goal of lattice QCD for several reasons:
\begin{itemize}
\item There are many `gold-plated' states: narrow, stable, and 
experimentally accessible.
\item The splittings in heavionium have particularly good properties for 
determining the 
lattice scale.
\item It is an important test of the actions used for $b$ and $c$, which 
can then be used to calculate decay constants which in turn are crucial 
for determining CKM matrix elements.
\item Control of systematic errors are well developed in calculations of
the $B$ and $D$ spectrum on the lattice due to relative insensitivity to
heavy-quark polarisation effects and to light quark masses.
\end{itemize}

 The Highly Improved Staggered Quark (HISQ) action \cite{Follana:2006rc,Follana:2007uv}
allows unprecedented control of discretization errors in 
numerical  lattice calculations. We use HISQ $s$ and $c$ valence quarks 
with NRQCD $b$ quarks on MILC lattices with $N_f=2+1$ flavors of ASQTAD 
sea quarks to calculate the masses of the $B_s$ and $B_c$ mesons.
\section{Heavy-light 2-point functions}
For increased statistics, we use random sources $\eta(x, t_0)$, 
defined as a 
three-component random complex unit-vector defined on each point in the 
source time slice $t_0$. These are the sources for the inversion of the 
HISQ strange and charmed valence quark propagators.  The HISQ
propagators, being staggered, are spinless and we need to convert them to
4-component propagators to combine with NRQCD propagators in a $b$-light
correlator. This is readily done at the sink by multiplying by the standard
staggered-to-naive transformation $\Omega(x) = \prod_i \gamma_i^{x_i}$. 
Since the
propagator source disappears once the propagator is made, the $\Omega$ factors
needed at the source (when a random wall is used) must be transferred to the
source of the heavy quark propagator and how to do this is described below.

NRQCD propagators are then made from a source which includes the same random
wall with which the HISQ propagators are made, plus the $\Omega$ factors and in
addition different Gaussian smearing factors
of varying radii $r_i$ chosen to allow improved overlap with the ground state
$B_s$ and $B_c$ mesons so that their energies can be extracted accurately at
early correlator times. We therefore initialise $N_{\rm smear}$ NRQCD $b$ 
propagators by setting 
\begin{equation}
G_i(x, t=0) = \sum_{x^\prime} S(\left|x-x^{\prime}\right|;r_i) \eta(x^\prime)  \Omega(x^\prime)
\end{equation}
and evolve with
\begin{equation}
G_i(x, t+1)= \left(1-\frac{\delta H}{2}\right)\left(1-\frac{H_0}{2n}\right)^nU_t^\dagger(x)\left(1-\frac{H_0}{2n}\right)^n\left(1-\frac{\delta H}{2}\right)G_i(x,t).
\end{equation}
We use an improved lattice NRQCD Hamiltonian \cite{Gray:2005ur}:
\begin{equation}
H_0=-\frac{\Delta^{(2)}}{2M^0}
\end{equation}
\begin{eqnarray}
\delta H &=& -c_1\frac{(\Delta ^{(2)})^2}{8(M^0)^3} + c_2\frac{ig}{8(M^0)^3}
({\Delta \cdot E - E \cdot \Delta})
-c_3\frac{ig}{8(M^0)^3}{\bf\sigma}\cdot({ {\Delta \times \tilde{E}} - {\tilde{E} \times \Delta}})\nonumber\\
&&-c_4\frac{g}{2M^0}{\bf\sigma}\cdot {\bf\tilde{B}} 
+ c_5\frac{a^2\Delta^{(4)}}{24M^0}
-c_6\frac{a(\delta^{(2)})^2}{16n(M^0)^2}.
\end{eqnarray}

Finally, at each time-slice, we combine the $N_{\rm smear}$ NRQCD propagators 
and the valence $s$ or $c$ quark propagator with the same smearing functions
at the sink end, giving us $N_{\rm smear}\times N_{\rm smear}$ $B_{s}$ 
and $B_{c}$ meson correlators.

\section{2-Point function effective masses and noise}
\begin{figure}
\begin{center}
\includegraphics[width=.6\textwidth]{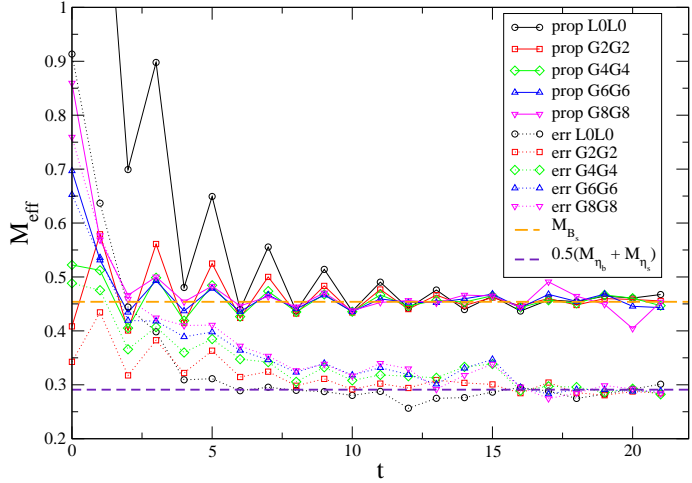}
\end{center}
\caption{Plot of effective masses of (local-local) 
$B_s$ correlator and $B_s$ correlator error. While the effective mass of the 
correlator matches the experimental $M_{B_s}$ (corrected for the energy shift),
 the effective mass of the correlator error gives 
$\frac{1}{2}(M_{\eta_b}+ M_{\eta_s})$. Several source-sink smearing 
combinations are shown. }
\label{eff_mass}
\end{figure}

The expression for the variance of the $B_s$ correlator 
\begin{equation}
\left[\langle G_{B_s}(i,j;t-t_0)G_{B_s}(i,j;t-t_0)\rangle - 
\langle G_{B_s}(i,j;t-t_0)\rangle^2\right]
\end{equation}
contains in it propagators for $\overline{b}\overline{s}bs$ four-quark states.
The lightest combination on the lattice is ${\eta_b}+{\eta_s}$. Therefore the 
error on the propagator falls like $e^{-\frac{1}{2}(M_{\eta_b}+ M_{\eta_s})t}$
while the signal falls like $e^{-M_{B_s}t}$, as 
can be seen in Figure \ref{eff_mass}. So the signal-to-noise ratio
degrades exponentially by the mass difference seen in the figure.
This degradation is worse than for $D_s$ states and necessitates smeared
sources to be able to fit propagators to small $t$ values.
Figure \ref{eff_mass} also illustrates that 
effective masses of propagators  $G_{B_s}(i,i;t-t_0)$ with
Gaussian smearing with radius 2 and 4 ($G2G2$ and $G4G4$) approach the $M_{B_s}$
plateau rapidly compared to the local source-sink combination ($L0L0$).

\section{Extracting $M_{B_s}$ and $M_{B_c}$}
We fit the measured $B_s$ and $B_c$ correlators to the form
\begin{equation} 
G_{\rm meson}(i,j;t-t_0)= \sum^{N_{\exp}}_{k=1} a_{i,k}a^*_{j,k}e^{-E_k(t-t_0)}
+ \sum^{N_{\exp}-1}_{k^\prime=1}   b_{i,k^\prime}b^*_{j,k^\prime}(-1)^{(t-t_0)}e^{-E^\prime_{k^\prime}(t-t_0)},
\end{equation} 
where $i$ and $j$ respectively index the source and sink smearing functions.
The second term is an oscillating parity partner state. 
We perform simultaneous Bayesian fits of the $G_{\rm meson}(i,j;t-t_0)$, looking 
for stability with respect to fit range and $N_{\exp}$.

Since the NRQCD Hamiltonian does not include a mass term, there is a shift in 
the energy of the $p=0$ states relative to the continuum mass. To correct for 
the energy shift in $M_{B_s}$ and $M_{B_c}$ we use the relationship:
\romeqn
\setcounter{equation}{0}
\begin{equation}
\label{EXPI}
M_{B_{s/c}} = \left(E_{B_{s/c}} -\frac{1}{2}E_{b\overline{b}}\right)_{\rm latt} + \frac{1}{2}M_{b\overline{b}},
\end{equation}
where $E_{B_s}$ or $E_{B_c}$ is the ground-state $E_1$.  
$M_{b\overline{b}}$ on the right-hand side is the
spin-averaged experimental masses of $b\overline{b}$ states:
$M_{b\overline{b}} = (3M_\Upsilon + M_{\eta_{b}})/4$, where we use the recent 
BaBar measurement of the $\Upsilon(1S)$-$\eta_b(1S)$ hyperfine 
splitting \cite{:2008vj}.
$E_{b\overline{b}}$ is the  corresponding spin-averaged lattice energy, 
calculated with NRQCD $b$ quark propagators on the same 
configurations \cite{Ikendall}.
For $M_{B_c}$ we also explore two other methods
for cancelling the energy shift:

\begin{equation}
\label{EXPII}
M_{B_c} = \left(E_{B_c} -\frac{1}{2}(E_{b\overline{b}}
+E_{c\overline{c}})\right)_{\rm latt} 
+ \frac{1}{2}\left(M_{b\overline{b}} + M_{c\overline{c}}\right),
\end{equation}
where $M_{c\overline{c}} = (3M_\psi + M_{\eta_{c}})/4$, and $E_{c\overline{c}}$ is 
the corresponding spin-averaged lattice energy.
The final method to extract $M_{B_c}$ is
\begin{equation}
\label{EXPIII}
M_{B_c} = \left(E_{B_c} - (E_{B_s}+E_{D_s} - E_{\eta_s})\right)_{\rm latt} 
+ \left(M_{B_s} + M_{D_s} - M_{\eta_s}\right),
\end{equation}
\reseteqn
where $M_{\eta_s}$ ($E_{\eta_s}$) is the mass (lattice energy) of a fictional
$s\overline{s}$ pseudoscalar: $M_{\eta_s} =\sqrt{2M_K^2 - M_\pi^2}$.

\begin{figure}
\begin{center}
\includegraphics[width=.6\textwidth]{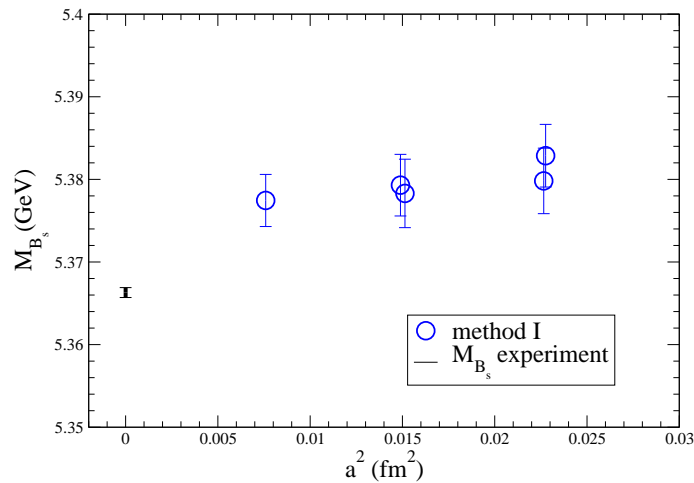}
\end{center}
\caption{Lattice calculations for $M_{B_s}$ on very 
coarse $16^3\times 48$, $a^{-1}\approx 1.3$ GeV, 
coarse $20^3\times 64$ and $28^3\times 64$, $a^{-1} \approx 1.6$ GeV
and fine $28^3\times 96$, $a^{-1} \approx 2.3$ GeV configurations.
}
\label{MBs_plot}
\end{figure}

\begin{figure}
\begin{center}
\includegraphics[width=.6\textwidth]{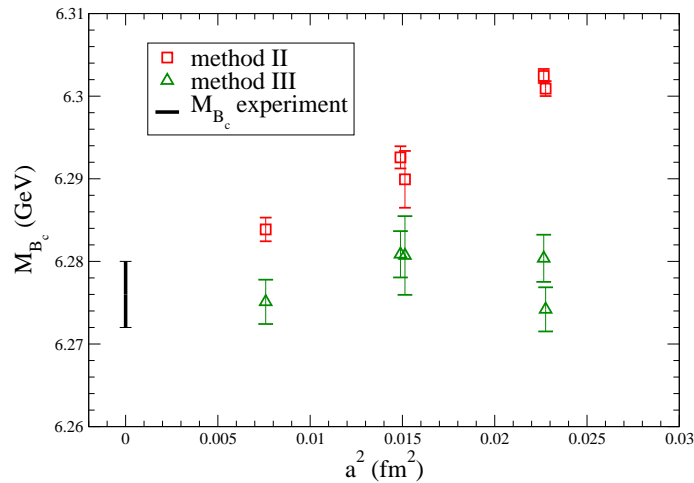}
\end{center}
\caption{Lattice calculations and experiment for $M_{B_c}$
on very coarse, coarse and fine configurations. For clarity we do not show
the calculations from method (I), with its significantly larger error 
bars.
}
\label{MBc_plot}
\end{figure}

We show results for several ensembles at three
lattice spacings and different light sea quark masses for
$M_{B_s}$ in Figure \ref{MBs_plot} and $M_{B_c}$ 
in Figure \ref{MBc_plot}. 
The statistical error (shown in figures) is dominated 
by the uncertainty in $a^{-1}$. We use MILC $r_1/a$ 
values \cite{Bernard:2001av,Aubin:2004wf} to set the scale 
ensemble-by-ensemble. We then convert to physical units with $r_1=0.321$fm
\cite{Gray:2005ur}.
An additional systematic error due to the $1.5\%$ uncertainty in 
$r_1$ must be added
in at the end with other systematics.
Expressions (\ref{EXPII}) and (\ref{EXPIII}) 
reduce the uncertainty from errors in $(r_1/a)$ by using  
$(E_{B_c} -\frac{1}{2}(E_{b\overline{b}}+E_{c\overline{c}}))_{\rm latt}$ and
$(E_{B_c} - (E_{B_s}+E_{D_s} - E_{\eta_s}))_{\rm latt}$, which are small
relative to the $\left(\right)_{\rm latt}$ quantity in (\ref{EXPI}).

\section{Conclusions}
While preliminary, these results show the potential for precise calculations
of $M_{B_s}$ and $M_{B_c}$ using HISQ valence quarks and NRQCD $b$ quarks on
a $2+1$ flavor ASQTAD sea. The lattice results show little sensitivity to
sea quark mass. Method (\ref{EXPI}) and to a lesser extent (\ref{EXPII})
for $M_{B_c}$ show a small 
dependence on the discretization scale.
A slightly mis-tuned $m_s$ in the coarse
and fine
ensembles \cite{Aubin:2004wf} may contribute some systematic 
error in $M_{B_c}$ values derived
from (\ref{EXPIII}), and will be corrected for in future work.
Relativistic corrections and electromagnetic 
effects are two possible sources of systematic error. Future efforts will 
improve statistics and include finer lattices.

\end{document}